\documentclass[12pt,preprint]{aastex}

\shorttitle{Variable Interstellar Absorption towards $\rho$ Leo}
\shortauthors{Lauroesch \& Meyer 2003}

\begin{document}


\title{Variable \ion{Na}{1} Absorption toward $\rho$ Leo : Biased
Neutral Formation in the Diffuse Interstellar Medium?\footnote{Based on
observations with the NASA/ESA {\it Hubble Space Telescope},
obtained from the data archive at the Space Telescope Science Institute.
STScI is operated by the Association of Universities for Research in
Astronomy, Inc. under NASA contract NAS 5-26555.}}


\author{J. T. Lauroesch\footnote{Visiting Astronomer, Kitt Peak National
Observatory (KPNO).  KPNO is a part of the National Optical Astronomy
Observatories (NOAO), which is operated by the Association of Universities
for Research in Astronomy , Inc.\ (AURA) under contract to the National
Science Foundation (NSF).} \ and David M. Meyer\footnotemark[2]}
\affil{Department of Physics and Astronomy, Northwestern University, Evanston,
	IL 60208}
\email{jtl@elvis.astro.northwestern.edu, davemeyer@northwestern.edu}


\begin{abstract}
We present multi-epoch KPNO Coud\'{e} Feed observations of interstellar
\ion{Na}{1} and \ion{Ca}{2} absorption toward the bright star $\rho$ Leo.
Comparisons of the \ion{Na}{1} profiles observed over a period of 8 years
reveal significant temporal variations in the \ion{Na}{1} column in at least
one component, implying that there is ``structure'' at scales of order of
the proper motion ($\sim$12 AU).  Archival {\it HST} Goddard
High Resolution Spectrograph observations of the \ion{C}{1} fine-structure
excitation in the variable component suggest that the density is
$\la$20 cm$^{-3}$, significantly lower than the densities inferred
in past \ion{H}{1} 21 cm and \ion{Na}{1} studies.  We suggest that the bulk
of the trace neutral species are in the density peaks within an interstellar
cloud.  The patchy distribution of these species naturally gives rise
to the large fluctuations seen on scales of 10--1,000 AU in past
temporal and binary studies.  This picture predicts that the
scales over which fluctuations will be observed vary as a function
of the ionization rate of a species.
\end{abstract}

\keywords{ISM: clouds -- ISM: structure -- stars: individual
	(HD~91316)}

\section{Introduction}

The sightline toward $\rho$ Leo (HD~91316) has long been a favorite of
spectroscopists, even before the interstellar origin of the so--called
``stationary'' lines was established.  \cite{harper1914} first noticed that
the \ion{Ca}{2} K line
toward $\rho$ Leo showed the characteristic difference in velocity
attributable to ``stationary'' lines, although the line was much wider than
was typically observed with hints that there were two blended lines present.
\cite{beals1936} showed that the \ion{Ca}{2} lines toward $\rho$ Leo were indeed
complex, making it one of the first three sightlines (along with $\epsilon$ and
$\zeta$ Ori) where multiple components were known.  Subsequently, \cite{smw1938}
showed that the \ion{Na}{1} lines also have multiple components, while
\cite{adams1943} showed that the ``violet'' \ion{Ca}{2} component of
\cite{beals1936} was itself multiple.  The velocities and equivalent widths
for the \ion{Na}{1} and \ion{Ca}{2} lines measured from photographic spectra
by \cite{rs1952} are (to within the errors) the same as those measured today.
Optical observations of $\rho$ Leo have continued to the present day with a
multitude of different instruments, and have shown the complex nature of the
\ion{Na}{1}, \ion{K}{1}, and \ion{Ca}{2} profiles \citep{hobbs1969, hobbs1971,
hobbs1974, whk94, wmh1996, wh2001}.  In addition, observations of numerous
ultraviolet absorption lines have been made by {\it Copernicus}, {\it IUE}, and
{\it HST}/Goddard High Resolution Spectrograph (GHRS). 

Over the past decade, high spatial resolution optical and radio studies
of the diffuse ISM have found strong evidence of pervasive subparsec-scale
variations in interstellar absorption lines \citep{l2001, c2003}, although
the precise interpretation of these fluctuations is not yet well understood
\citep{heiles97, bge99, deshpande2000, fg2001, wf2001, l2001, c2003}.
Initially, these observations were interpreted as ubiquitous concentrations
of apparently dense atomic gas ($n_H\gtrsim$ 10$^3$ cm$^{-3}$) in otherwise
diffuse sight-lines.  However, recent direct measurements of the densities in
a few of the optically selected clouds have suggested that in general the
densities are much lower than previously inferred \citep{lmwb98, lmb2000,
wf2001}, although observations in clouds identified by tracers of dense gas
(\ion{K}{1}) suggest that at least some of these features are associated
with denser structures \citep{pfw2001, c2002}.  Thus, it has been suggested that
small-scale variations in the physical parameters may be responsible for
much (if not all) of the observed variation in optical studies of
small-scale structure.  Recently, as part of a larger survey to
identify very small scale (of order 10 AU) variations in interstellar lines,
evidence for temporal (proper motion induced) variations in the interstellar
\ion{Na}{1} line toward $\rho$ Leo have been observed.   In this Letter, we
will discuss these new observations as well as our analysis of archival
Goddard High Resolution Spectrograph (GHRS) echelle observations of this
sightline, and then discuss a simple model of neutral formation in
the context of these results.

\section{Observations and Data Reduction}

Rho Leo is a bright (V=3.8) B1Ib runaway star, located at a distance of
about 870 pc approximately 690 pc above the galactic plane \citep{ds94}.
The proper motion is 6.6$\pm$0.9 mas/yr \citep{perryman1997}, which
corresponds to a projected motion of 1.56$\pm$0.21 AU/yr.
Since 1989, we have obtained multiple observations of the interstellar
\ion{Na}{1} and \ion{Ca}{2} absorption lines toward $\rho$ Leo using the
KPNO Coud\'{e} Feed and spectrograph at resolutions of $\sim$1.4 and
3 km s$^{-1}$ (see Table~1).  These datasets have been reduced using the NOAO
IRAF\footnote{IRAF is distributed by NOAO, which are operated by AURA, Inc.\
under cooperative agreement with the NSF.} data reduction packages in the
same fashion as that used by \cite{wm96} and \cite{lm99}.  The extracted
continuum--fitted spectra for all observing runs are shown in Figure~1 -- note
in particular the variation in the \ion{Na}{1} profile at a heliocentric
velocity of 18 km s$^{-1}$.

\smallskip
\centerline{Table 1: $\rho$ Leo Observation Log}
\begin{table}[h]
\begin{center}
\begin{tabular}{lccc}
\tableline\tableline
Species & Date & Resolution & S/N ratio \\
        &      & (km s$^{-1}$) & (per pixel) \\
\tableline
\ion{Na}{1} & 11/95 & 1.4 & 370 \\
            & 11/01 & 3.4  & 200 \\
            & 11/02 & 3.4  & 220 \\
            & 02/03 & 1.4 & 250 \\
\ion{Ca}{2} & 05/89 & 4.8 & 190 \\
            & 01/92 & 3.6 & 400 \\
            & 11/02 & 3.6 & 330 \\
\ion{Al}{3}\tablenotemark{a} & 12/1992 & 3.6 & 35--55 \\
\ion{C}{1}\tablenotemark{a,b} & 03/1996 & 3.5 & 70 \\
\ion{Zn}{2}\tablenotemark{a} & 12/1992 & 3.4--3.6 & 70--80 \\ \tableline
\multicolumn{4}{l}{\tablenotemark{a}\ \ Data obtained from the {\it HST}
	archive, obtained} \\
\multicolumn{4}{l}{\ \ as parts of programs 2251 and 5882 (PI Hobbs).}\\
\multicolumn{4}{l}{
\tablenotemark{b}\ \ Observations of the $\lambda\lambda$ 1328\AA\ multiplet.}\\
\end{tabular}
\end{center}
\end{table}

\begin{figure}
\begin{center}
\epsscale{.7}
\plotone{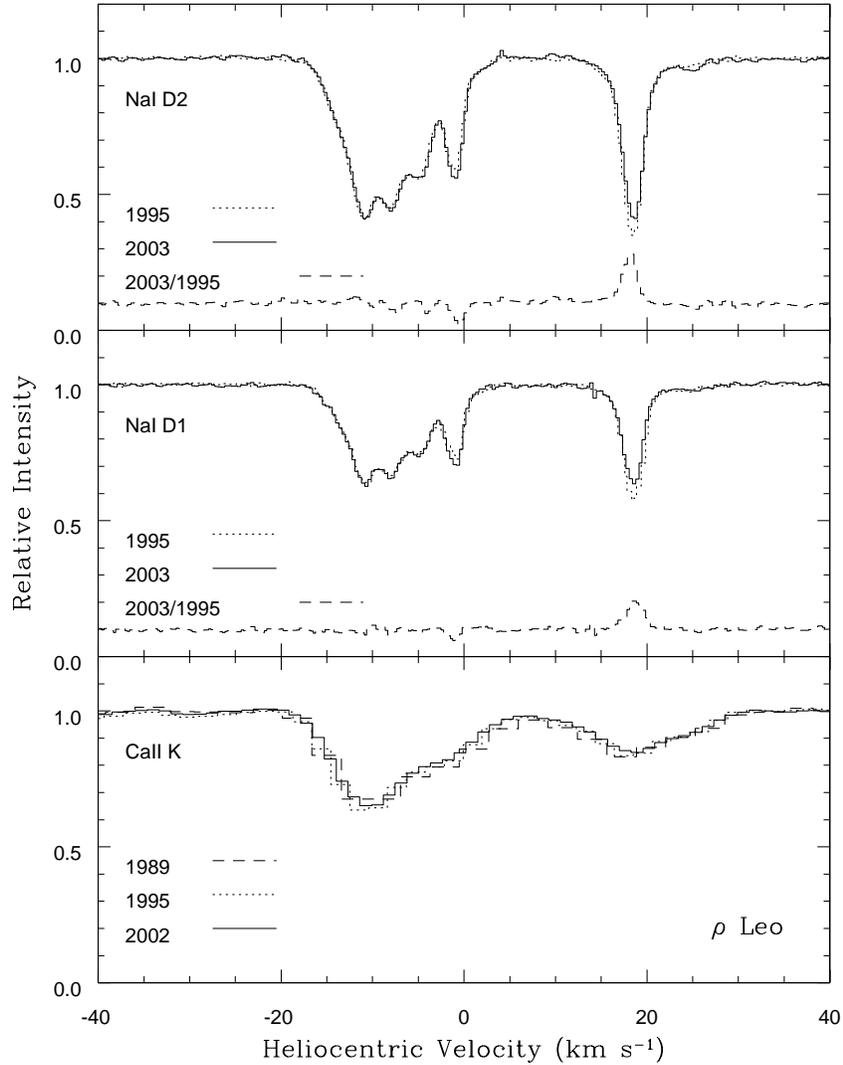}
\end{center}
\figcaption[rholeo_reobs2.ps]{Observations of the \ion{Na}{1} D$_2$ (5889\AA),
\ion{Na}{1} D$_1$ (5895\AA), and \ion{Ca}{2} K (3933\AA) lines towards
$\rho$ Leo.  The top panel shows our two high resolution (1.4 km s$^{-1}$)
observations, note in particular the variation in the \ion{Na}{1} profile near
18 km s$^{-1}$, as well as the somewhat more subtle variation near
-1 km s$^{-1}$.  Also shown at the bottom of the plot is the 2003 profile
divided by the 1995 profile (with an offset of -0.9 applied to the relative
intensity for plotting purposes).  The lower panel shows our three moderate
resolution observations of the \ion{Ca}{2} K line.}
\end{figure}

In addition to the ground--based data, we have taken advantage of
existing archival GHRS spectra of $\rho$ Leo.  Observations using multiple
settings of the ECH-A and ECH-B gratings taken in March 1996 and December
1992 (respectively) are available from {\it HST} observing programs 5882
and 2251 (PI Hobbs).  These datasets were reduced
using the standard STSDAS routines (POFFSETS, DOPOFF, and SPECALIGN) for both
the WSCAN (ECH-B) and FP-SPLIT (ECH-A) datasets, and then selected
transitions were continuum fitted and analyzed (Figure~2).

\begin{figure}
\plotone{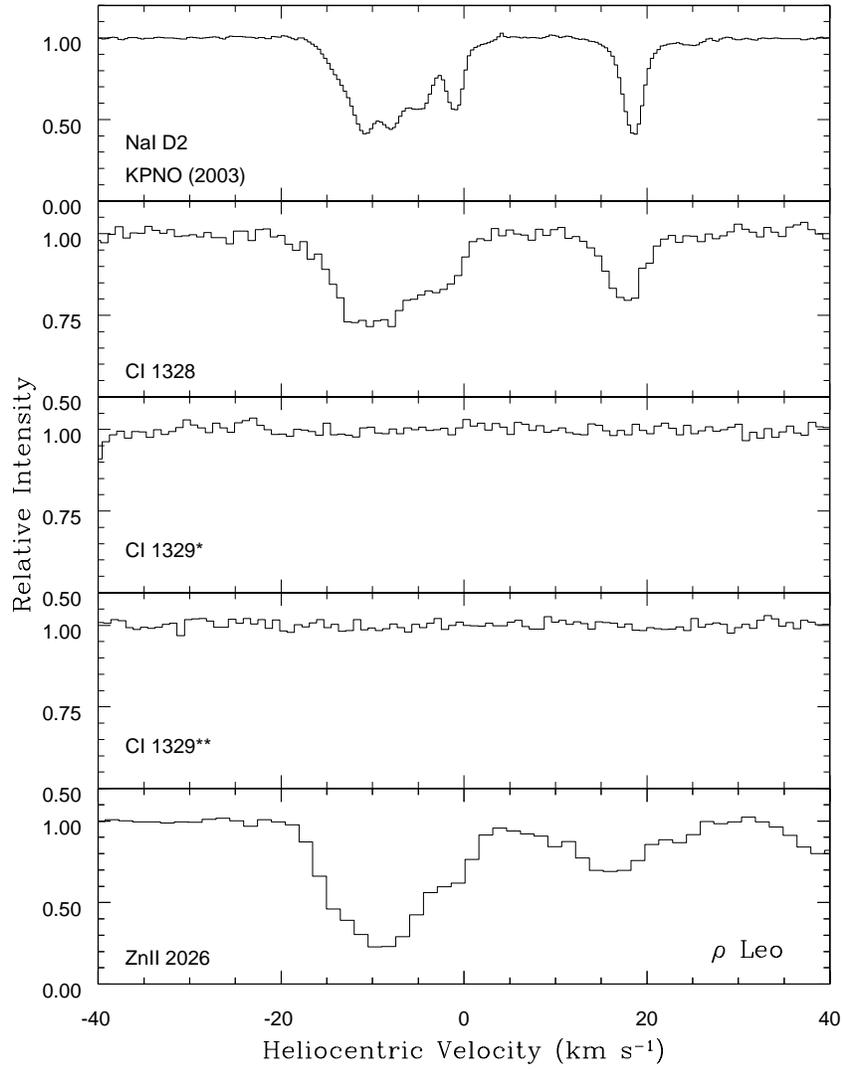}
\figcaption[rholeo_hst.ps]{The \ion{C}{1} 1328 multiplet and the \ion{Zn}{2}
2026 line toward $\rho$ Leo, with the 2003 \ion{Na}{1} observation shown
for scale.  Note the lack of excited \ion{C}{1}$^{\star}$ and
\ion{C}{1}$^{\star\star}$ absorption -- the plots are centered on the
strongest transitions at 1329.1004\AA\ and 1329.5775\AA\ respectively.}
\end{figure}

The column densities, b--values, and relative velocities of the various
components in both the KPNO and GHRS datasets were derived by profile fitting
the continuum fitted spectra using the program {\tt fits6p} \citep{why91}.
The wavelengths and oscillator strengths were taken from \cite{morton91}
and \cite{welty99}, with the hyperfine splitting of the \ion{Na}{1} D lines
taken into account \citep{whk94}.  The \ion{Na}{1} component model of
\cite{whk94} was used as a starting point for fitting the neutral species,
with additional input from the \ion{Ca}{2} component model of \cite{wmh1996}
used for fitting the ionized species.  The column density, b-value, and
velocity of the various components were allowed to be free parameters in
the initial fits -- in the final fits the relative velocities of the
components were fixed and the variation in the b-value limited for
most components.

\section{The v$\sim$18 km s$^{-1}$ Cloud}

As noted above, the most striking feature when comparing the \ion{Na}{1}
profiles in Figure~1 is the strong temporal variation near 18 km s$^{-1}$.
Also significant, however, is the lack of a
corresponding variation in the observed \ion{Ca}{2} profiles, as well as the
lack of convincing \ion{Na}{1} profile variations at other velocities.
We should note here that part of the reason for including $\rho$ Leo in our
survey was the apparent lack of \ion{H}{1} 21-cm emission associated with
the 18 km s$^{-1}$ component in the survey of \cite{hobbs1971}, although
more recent surveys show some 21-cm emission at this velocity in the
region around $\rho$ Leo \citep{hb1997}.  In addition to showing clear
evidence for temporal variability, the v$\sim$18 km s$^{-1}$ component is also
relatively isolated and so we will concentrate on this component for the
rest of the discussion.

The column densities for selected species derived for the
v$\sim$18 km s$^{-1}$ component are given in Table~2.  The first thing
to note is the large \ion{Na}{1} differences between the years 1989 and 2003,
corresponding to a decrease of almost one--third in the column.  However,
there are two important caveats -- first, we assumed the same b-value
(0.58 km s$^{-1}$) for the component as that found by \cite{whk94} using
their higher resolution data.  Second, and perhaps more important, the
published \cite{whk94} profile shows no sign of the weak absorption redward
of this component seen in our datasets.  We note that this shelf of absorption
is seen in the profile published by \cite{hobbs1971, hobbs1974} which has
a similar resolution to our datasets.  This suggests that these features may
have been over-resolved in the \cite{whk94} data and ``lost'' during continuum
placement, perhaps altering the derived column density.  In any case,
as shown in Figure~1, there are significant changes in the \ion{Na}{1} column
even over the past 8 years, with a decrease of $\sim$20\%\ if we assume the
\cite{whk94} b-value.  A period of 8 years corresponds to a projected
proper motion at the distance to the star of $\sim$ 12 AU.
One useful comparison that can be made immediately
is to compare the N(H) one would estimate from the observed \ion{Na}{1}
columns, to that which would be derived using the observed N(\ion{Zn}{2}).
The average of the 1989 and 1995 \ion{Na}{1}
columns is 27.5$\times 10^{10}$ cm$^{-2}$, from which one would infer an N(H)
of $\sim$5.5$\times 10^{19}$ cm$^{-2}$ \citep{ferlet85}.  Similarly, one would
estimate N(H)$\sim$4.6--7.2$\times 10^{19}$ cm$^{-2}$ from N(\ion{Zn}{2})
(assuming a Zn depletion of 0.2 to 0.4 dex \citep{welty99}).
This consistency immediately highlights a potentially important difference
between this cloud and the clouds studied previously in the ultraviolet by
\cite{lmwb98} and \cite{lmb2000}, namely the observed N(\ion{Zn}{2}) (and
inferred N(H)) is approximately an order of magnitude larger for this cloud,
even though the N(\ion{Na}{1}) is only a factor of 2--3 larger.

\smallskip
\centerline{Table 2: Selected Column Densities at v=18 km s$^{-1}$ toward
        $\rho$ Leo}
\begin{table}[h]
\begin{center}
\begin{tabular}{lccc}
\tableline\tableline
Species & Year & Lines/Multiplet & N(X)\\
        &      & Used (\AA)      & (cm$^{-2}$) \\
\tableline\tableline
\ion{Na}{1} & 1989\tablenotemark{a} & 5895 & 30.7$\times 10^{10}$ \\
         & 1995 & 5889,5895& 24.3$\pm$0.5$\times 10^{10}$\ \tablenotemark{b} \\
         & 2001 & 5889,5895& 20.8$\pm$1.0$\times 10^{10}$\ \tablenotemark{b} \\
         & 2002 & 5889,5895& 20.2$\pm$1.0$\times 10^{10}$\ \tablenotemark{b} \\
         & 2003 & 5889,5895& 20.9$\pm$0.6$\times 10^{10}$\ \tablenotemark{b} \\
\ion{Al}{3} & 1992 & 1854, 1862 & 5.9$\pm$1.1$\times 10^{11}$ \\
\ion{C}{1} & 1996 & 1328 & 57.7$\pm$6.2$\times 10^{11}$ \\
\ion{C}{1}$^{\star}$ & 1996 & 1328 & $<$8.4$\times 10^{11}$\ \tablenotemark{d}\\
\ion{C}{1}$^{\star\star}$ & 1996 & 1328 & $<$8.6$\times 10^{11}$\ \tablenotemark{d}\\
\ion{Cr}{2} & 1992 & 2056, 2062 & 1.8$\pm$0.2$\times 10^{11}$ \\
\ion{Mg}{1} & 1992 & 2026 & 17.6$\pm$1.8$\times 10^{11}$ \\
\ion{Zn}{2} & 1992 & 2026, 2062 & 13.1$\pm$1.8$\times 10^{11}$ \\
\tableline
\end{tabular}
\end{center}
\end{table}

We can now use the other species listed in Table~2 to estimate the
physical conditions in this cloud.  We can use the observed
N(\ion{C}{1}) and the upper limit on N(\ion{C}{1}$\star$) to place
limits on the pressure and density in this cloud \citep{js79} --
our upper limits to N(\ion{C}{1}$\star$) imply for
T$\sim$100--500 K (consistent with the observed b-value)
pressures of log(p/k)$<$3.3--3.8 cm$^{-3}$ K.  For an ideal gas at these
temperatures, such pressures correspond to densities of n(H)$\la$20 cm$^{-3}$,
similar to the densities found in previous studies \citep{lmwb98, lmb2000}.
In addition, the fine--structure equilibrium can also be used to estimate
the minimum distance between the cloud and $\rho$ Leo itself, since
photon--pumping can significantly alter the observed level populations
\citep{js79, lmwb98}.  The observed upper limits to N(\ion{C}{1}$\star$)
suggest that this cloud is $\ga$10 pc away from $\rho$ Leo.  The upper limit
on the density coupled with our estimates of N(H) above suggest a cloud size
of order 1 pc, much larger than the length scale over which we are seeing
\ion{Na}{1} fluctuations ($\sim$12 AU) but also smaller than the
estimated distance between the cloud and $\rho$ Leo.
The observed ratio of N(\ion{Cr}{2})/N(\ion{Zn}{2})$\sim$1.4 would suggest
that this cloud has a ``warm cloud'' depletion pattern \citep{welty99}.  If
we now assume a typical ``warm cloud'' value for the depletion of Al, we would
estimate N(\ion{Al}{2}$+$\ion{Al}{3})$\sim$1.1$\times 10^{13}$ cm$^{-2}$ from
N(\ion{Zn}{2}), which would suggest that while \ion{Al}{2} is the dominant
ionization state in this cloud there is likely to be some degree of
photo-ionization.  We can also use the measured neutral and \ion{Zn}{2} column
densities to estimate electron densities of 0.04, 0.15, and 0.05 cm$^{-3}$
from \ion{C}{1}, \ion{Na}{1} and \ion{Mg}{1} respectively assuming T$=$100 K
and a `warm cloud'' depletion pattern \citep{pa86, lmb2000}.  If we assume the
dominant source of electrons is singly ionized carbon, these values correspond
to a density of $\sim$100--400 cm$^{-3}$ at T$=$100 K, significantly in
excess of the upper limit derived above.

\section{Biased Neutral Formation in Interstellar Gas?}

The parameters derived above for the v$\sim$18 km s$^{-1}$ component toward
$\rho$ Leo appear fairly typical for a diffuse cloud.  The column density
of hydrogen is $\sim$6$\times 10^{19}$ cm$^{-2}$, and we would estimate a
cloud size of $\sim$1 pc along the line of sight.  The density of hydrogen
appears to be of order 10 cm$^{-3}$, with the derived electron density and
observed \ion{Al}{3} suggesting some partial ionization of hydrogen.  There is
one significant problem that still remains however, namely the origin of
the observed fluctuations in the \ion{Na}{1} profile over relatively short
timescales.  Since the photo-ionization and recombination rates for
\ion{Na}{1} are relatively low \citep{pa86}, one would not expect to see
significant variations in the column on timescales shorter than thousands of
years in an interstellar cloud.  If we were thus to assume the observed
fluctuations set the transverse size of the cloud, the earliest observations
of $\rho$ Leo suggest that significant \ion{Na}{1} and \ion{Ca}{2} absorption
has been detectable at this velocity for at least 65 years, setting a
minimum transverse size of 100 AU for this ``structure'' (at the distance
of the star).  The result is similar to the models suggested by
\cite{heiles97}, namely a very long, thin sheet or filament with
a thickness of order one thousandth the length.  However, there is
an alternative explanation for this observed variation.

Consider an isolated diffuse interstellar cloud -- it will have within
it density fluctuations at scales much smaller than a parsec \citep{bge99,
clv2002}.  Just as the force of gravity biases galaxies to form in the peaks
of the cosmic density distribution, the neutral species in interstellar clouds
are biased by the recombination rates to form on the peaks of the density
distribution within a cloud.  It is well known that recombination rates have
a strong density dependence \citep{pa86}, indeed this formed the basis for
previous suggestions that relatively small density fluctuations could give
rise to the observed small scale structure \citep{lm99, lmb2000}.
Furthermore, the \ion{C}{1} survey of \cite{jt2001} has identified a
population of high pressure components in the ISM, which could represent
these density peaks.   We now ask what are some of the observational
consequences of such a picture?

First we can estimate a minimum apparent length scale for the size of
fluctuations observable in any trace species.  The scale is just set by
the distance a neutral can travel before it can be ionized, which in low
density gas is of order the velocity of an atom divided by the
photo--ionization rate.  Assuming v$\sim$0.5 km s$^{-1}$ which is a
``typical'' b-value for the known temporally variable components,
we derive an apparent scale of order 250 AU for \ion{Na}{1} structures.
Despite our having neglected to correct for collisions, this size is of
order that observed in temporal fluctuations to date, and is consistent with
the observations of \ion{Na}{1} towards binary star systems.  Furthermore, note
that this scale is dependent upon the photo-ionization rate for a given species,
and thus (for example) for \ion{Ca}{2} we would obtain a scale about 5 times
larger.  This result very simply explains why we do not observe a corresponding
variation in the \ion{Ca}{2} profile towards $\rho$ Leo.  In addition, it
suggests that species with higher photo-ionization rates such as \ion{Ca}{1}
and (perhaps) \ion{C}{1} should show larger column density variations than
species such as \ion{Na}{1} or \ion{K}{1}.  Indeed, \cite{c2002} used
observations of \ion{Ca}{1} to suggest that the variable component towards
$\kappa$ Velorum arises in very dense material ($\rm n_H\ga 10^{3} cm^{-3}$).

A further consequence of this picture is that only a fraction of the
total gas would reside in these peaks.  Thus, adjacent sightlines should show
smaller changes in the columns of the dominant ions, since the size of the
peaks is only a fraction of the size of a cloud the additional column density
for the dominant species would be significantly reduced.  Consequently,
the neutral fraction within the density peaks would be higher (perhaps much
higher) than what one would calculate by taking the integrated columns,
suggesting that the electron densities inferred from ionization balances are
too low.  Furthermore, the fine-structure equilibrium of \ion{C}{2} and
\ion{Si}{2} would also be affected.  Just like the neutral species, most of
the observed \ion{C}{2}$^{\star}$ and \ion{Si}{2}$^{\star}$ absorption would
arise in the density peaks, but the columns of the ground-states would
reflect the large scale distribution of gas and would include the
perhaps dominant contribution by the material outside the peaks.  Thus,
in this picture, these methods provide only an ill--defined ``average''
electron density within a cloud, and would explain why there is so much
difficulty in deriving self-consistent electron densities in many cases
\citep{welty99}.  Assumptions about partial ionization of hydrogen could
then not be made using the (separately) derived electron and total
hydrogen densities, since these tracers will likely be probing different
portions of a cloud. 

A final question is whether there is a connection between the
observed small scale variations in \ion{Na}{1} and the high pressure
clumps observed by \cite{jt2001}.  In this picture, the fine--structure
equilibrium of \ion{C}{1} becomes a measure of some sort of ``average''
density of the peaks along a sightline through a cloud.  To date, there
have been limited observations of the \ion{C}{1} fine structure equilibrium
associated with optically selected small scale structures
\citep{lmwb98, lmb2000, wf2001}.  Including $\rho$ Leo, the observed
\ion{Na}{1} structures in all of these sightlines appear to arise in
relatively low density gas, and not in the high pressure clumps detected
by \cite{jt2001}.  However, as \cite{c2003} recently noted for the temporally
variable components towards both HD~32040 and HD~219188, these sightlines
may be sampling a mixture of low pressure gas with a fraction (perhaps 20\%)
of high pressure, dense material with which the temporal variation may be
associated.  In the picture developed above, the observed temporal and
binary star variations are arising in the neutral ``halo'' around one or more
of these high pressure clumps, implying that these clumps are a fraction
of the apparent size we calculated above.  Yet another possibility is that
we are seeing the remnant neutrals formed in transient high pressure clumps
which have already expanded into the ambient cloud.  Planned re--observations
of both HD~32040 and HD~219188 will likely help in beginning to resolve this
question.   In any case, this suggests that the density of the majority of the
gas in a cloud is smaller than that inferred from the \ion{C}{1} observations,
but that some fraction resides in higher density structures within which
arises the bulk of the trace neutral species.

\acknowledgments
The authors thank the referee Ian Crawford for his valuable comments
on this manuscript.  It is also a pleasure to acknowledge
the support of the staff of KPNO, especially Daryl Willmarth, for their
assistance in obtaining this data.

\end{document}